# Low-energy electron scattering from atomic Th, Pa, U, Np and Pu: Negative ion formation

**Zineb Felfli and Alfred Z. Msezane**
Department of Physics and Center for Theoretical Studies of Physical Systems, Clark Atlanta University, Atlanta, Georgia 30314, USA

**Abstract**

Here we investigate ground and metastable negative ion formation in low-energy electron collisions with the actinide atoms Th, Pa, U, Np and Pu through the elastic total cross sections (TCSs) calculations. For these atoms, the presence of two or more open d- and f- subshell electrons presents a formidable computational task for conventional theoretical methods, making it difficult to interpret the calculated results. Our robust Regge pole methodology which embeds the crucial electron correlations and the vital core-polarization interaction is used for the calculations. These are the major physical effects mostly responsible for stable negative ion formation in low-energy electron scattering from complex heavy systems. We find that the TCSs are characterized generally by Ramsauer-Townsend minima, shape resonances and dramatically sharp resonances manifesting ground and metastable negative ion formation during the collisions. The extracted from the ground states TCSs anionic binding energies (BEs) are found to be 3.09 eV, 2.98 eV, 3.03 eV, 3.06 eV and 3.25 eV for Th, Pa, U, Np and Pu, respectively. Interestingly, an additional polarization-induced metastable TCS with anionic BE value of 1.22 eV is created in Pu due to the size effect. We also found that our excited states anionic BEs for several of these atoms compare well with the existing theoretical electron affinities including those calculated using the relativistic configuration-interaction method. We conclude that the existing theoretical calculations tend to identify incorrectly the BEs of the resultant excited anionic states with the electron affinities of the investigated actinide atoms; this suggests a need for an unambiguous definition of electron affinity.

**Keywords:** generalized bound states; electron correlations; anionic binding energies; Regge poles; actinide atoms

1. Introduction

One of the most challenging problems in atomic and molecular physics, when exploring negative ion formation in complex heavy atoms and fullerene molecules and to date still continues to plague both experiments and theory, is the determination of accurate and reliable values for the important electron affinity (EA) of the atoms and molecules involved. Accurate and reliable atomic and molecular affinities are essential for understanding chemical reactions involving negative ions [1], whose importance and vast utility in terrestrial and stellar atmospheres as well as in device fabrication, drug delivery and organic solar cells are well-documented. In the lanthanide and actinide atoms the presence of two or more open d- and f- subshell electrons results in enormous numbers of complicated and diverse electron configurations that characterize low-energy electron interactions in these systems. These present formidable computational complexity when using conventional theoretical methods that renders obtaining accurate and reliable EAs very difficult, if not impossible. Thus the published literature abounds in incorrect EAs for the lanthanide and actinide atoms. These reflect the difficulties in the theoretical understanding of the fundamental mechanism responsible for low-energy electron attachment in these systems leading to stable negative ion formation. Incidentally, the binding energy (BE) of the least-bound electron in nobelium, corresponding to the first ionization potential, has recently been determined experimentally with high precision [2]. However, to our knowledge no such experimental determination of the EAs of the actinide atoms has been reported.

While the generation of the popular but notoriously slow-converging large partial wave expansions offers no physical insights, the complex angular momentum (CAM) methods have the advantage in that the calculations are based on a rigorous definition of resonances, viz. as singularities of the S-matrix. Recently, a theoretical breakthrough was achieved in low-energy electron scattering from complex heavy atoms and fullerene molecules through the use of the Regge pole, also known as the CAM methodology. The crucial electron-electron correlation effects and the vital core-polarization interaction have been identified as the major physical effects mostly responsible for electron attachment in low-energy electron scattering from complex heavy atoms and fullerene molecules, leading to stable negative ion formation. Indeed, the low-energy electron elastic scattering total cross sections (TCSs) for these systems have been found to be characterized generally by Ramsauer-Townsend (R-T) minima, shape resonances (SRs) and dramatically sharp resonances manifesting ground and metastable anionic formation during the collisions. For the fullerene molecules the extracted from the TCSs ground state anionic binding energies matched excellently the measured EAs for the fullerene molecules from $C_{20}$ through $C_{92}$ [3, 4] for the first time. This is an unprecedented theoretical feat that has never been achieved before and the Regge pole approach requires no assistance whatsoever from either experiment or other theory to achieve the remarkable results.

Interest in the investigation of low-energy electron elastic scattering from the actinide atoms Th, Pa, U, Np and Pu as well as the already studied fullerene molecules $C_n$ ($n$=20,------112) [3, 4] leading to ground and metastable negative ions formation is stimulated by the following. Electronic structure and stabilization of $C_{60}$ through the encapsulation of actinide atoms An@$C_{60}$ (An = Th, ------, Md) has been investigated theoretically [5]. The study of the charged and neutral endohedral fullerenes An@$C_{28}$ (An = Th,-------, Md) [6, 7] and An@$C_{40}$ (An = Th,-------, Md) [8] concluded that some of the clusters could be very stable. Therefore, they could be useful in nuclear applications such as in medicine and/or nuclear waste disposal. Notably, the U@$C_{60}$ is the only actinide endohedral system that has been measured in arc-produced soot [9]. The experiment [10] concluded that the encapsulated metal species, M@$C_{28}$ (M= Ti, Zr, U) may catalyze or nucleate endohedral fullerene formation. Indeed, the M@$C_{60}$ fullerene hybrids have demonstrated catalytic efficiency in fundamental hydrogenation [11]. For the lanthanide endohedral systems Gd@$C_n$ ($n$ = 74, 76, 78, 82, 84), the EAs were measured and found to remain close to those of the empty fullerenes [12]. Clearly in this context the understanding of low-energy electron scattering from fullerene molecules and complex heavy atoms, leading to negative ion formation is essential, including the determination of the theoretically challenging calculation of the corresponding EAs of these systems. Consequently, the low-energy electron scattering from the actinide atoms Th, Pa, U, Np and Pu is explored here through the elastic TCSs calculations primarily to identify and delineate the resonance structures as well as extract the anionic ground state BEs which are identified with the EAs of these actinide atoms.

In our continuing investigations of low-energy electron scattering cross sections for complex heavy atoms and fullerene molecules, the identification of the ground and metastable anionic formation is essential for understanding the fundamental mechanism of negative ion formation. Appropriately, our robust Regge pole methodology is employed for the TCSs calculations, since Regge poles are generalized bound states. Very recently, the ground state anionic binding energies extracted from our Regge pole calculated electron elastic TCSs for the fullerenes $C_{20}$ through $C_{92}$ have been found to match excellently the measured EAs [3, 4]. While for the complex heavy atoms Au and Pt the agreement between our Regge pole calculated ground state anionic BEs and the measured EAs of Au [13-15] and Pt[16] is outstanding, for the Eu, Tm and other lanthanide atoms, including Nb the resultant conundrum between our calculated ground state anionic BEs and the measured EAs has been discussed, with the conclusion that the measured EAs for the atoms Tm and Tb, require reinterpretation [17], including the recently measured EAs for Eu [18] and Nb [19]. For the lanthanide atoms Eu, Tm and Tb as well as others, including Nb, the presence in the calculated TCSs of both ground and metastable anionic resonances led to the misidentification of the measured EAs, see Ref. [17].

The unprecedented success of our Regge pole methodology in the calculation of low-energy electron elastic scattering TCSs for complex heavy atoms and fullerene molecules is due mainly to the fact that embedded in our Regge pole methodology are the vital electron-electron correlation effects and the crucial core-polarization interaction. Indeed, for the understanding and correct interpretation of confinement resonances in the photoionization of the endohedral atom $A@C_n$ ($C_n$ = fullerene) the identification and delineation of the resonances in the TCSs of the isolated atom A and the fullerene $C_n$ are essential. In [3, 4] the presence of the ground and the metastable negative ion formation in low-energy electron scattering from many fullerenes ($C_{20}$ to $C_{240}$) have already been investigated through the TCSs calculations. For the actinide atoms Pa and U it has been concluded recently from preliminary results [20] that the existing calculations incorrectly identified the BEs of their excited anions with the EAs [21, 22, 23]. Here our calculated ground and metastable states TCSs for the very complicated structurally actinide atoms Th, Pa, U, Np and Pu will demonstrate and clarify the assertion. Indeed, the obtained results here demonstrate unequivocally the need to identify and delineate the resonance structures in the low-energy electron scattering from complex heavy atoms and fullerene molecules, with the objective to identify precisely the anionic ground state BEs which yield the desired atomic or molecular EAs. The electron attachment process in complex heavy systems is generally characterized by stable ground, metastable and excited negative ion formation; hence the need for the identification of their BEs, particularly those corresponding to the ground state because they yield directly the desired EAs.

The structure of the paper is as follows. Section 2 presents briefly the method of calculation, while Sections 3 and 4 deal with the Results and the Discussion and Conclusion, respectively.

## 2. Method of Calculation

Most existing theoretical methods used for calculating the anionic BEs of complex heavy systems, including those for the $Au^-$ and the $C_{60}^-$ anions, are structure-based. Therefore the results obtained through these methods are often riddled with uncertainties and lack definitiveness for complex heavy systems. Their general weakness is that they employ large configuration interaction expansions demanded by the diffuseness of the wave functions for negative ions. The resultant computational complexity limits the theoretical understanding of the fundamental mechanism underlying low-energy electron collisions with complex heavy systems leading to stable negative ion formation. Regge poles are generalized bound states; therefore if the essential physics is accounted for adequately as in our case, the anionic BEs of the ground state complex heavy systems will be calculated reliably [3, 4, 17].

Our Regge pole methodology provides a novel and robust theoretical approach to the seemingly impossible task of determining definitively stable ground and metastable negative ion formation in low-energy electron collisions with complex heavy atoms such as the actinide atoms and the fullerene molecules. The approach avoids entirely the complications with the attendant uncertainties encountered by conventional theoretical methods by considering the incident electron to interact with the actinide atom without the consideration of the complicated details of the electronic structure of the actinide atom itself. The Regge pole methodology requires no assistance whatsoever from either experiment or other theory to achieve the impressive results presented in this paper and elsewhere [3, 4].

## 2.1 Elastic Cross Section

The Regge pole, also known as the complex angular momentum (CAM) methodology is appropriate for investigating low-energy electron scattering from the complex heavy atoms/fullerene molecules, resulting in negative ion formation as resonances since Regge poles, singularities of the S-matrix, rigorously define resonances [24, 25]. Embedded in our Regge pole methodology are the crucial electron-electron correlations and the vital core-polarization interaction. These effects are mostly responsible for the existence and stability of typical negative ions. The fundamental quantities which appear in the CAM theories are the energy-dependent Regge pole positions, $\lambda_n$ and the corresponding residues, $\rho_n$ where $n = 0, 1, 2, ...$; they determine the cross sections, both differential and integral. Plotting Im $\lambda_n$ (E) versus Re $\lambda_n$ (E) ($\lambda_n$ (E) is the CAM) the well-known and revealing Regge trajectories can be investigated [26]; they probe electron attachment at the fundamental level near threshold. Their importance in low-energy electron scattering has been demonstrated in for example [26, 27].

Within the CAM representation of scattering, the Mulholland formula [28] for the electron elastic TCS takes the form [29, 30] (atomic units are used throughout):

$$\sigma_{tot}(E) = 4\pi k^{-2} \int_0^\infty \mathrm{Re}[1 - S(\lambda)]\lambda d\lambda$$
$$- 8\pi^2 k^{-2} \sum_n \mathrm{Im} \frac{\lambda_n \rho_n}{1+\exp(-2\pi i \lambda_n)} + I(E) \quad (1)$$

In Eq. (1) S is the S-matrix, $k = \sqrt{2mE}$, with $m$ being the mass and $E$ the impact energy, $\rho_n$ is the residue of the S-matrix at the $n^{th}$ pole, $\lambda_n$ and $I(E)$ contains the contributions from the integrals along the imaginary $\lambda$-axis; its contribution has been demonstrated to be negligible [26]. Here we consider the case for which Im $\lambda_n \ll 1$ so that for constructive addition Re $\lambda_n \approx 1/2, 3/2, 5/2...$, yielding $l = \mathrm{Re}\, \lambda_n \cong 0, 1, 2....$ The significance of Eq. (1) is that a resonance is likely to influence the elastic TCS when its Regge pole position is close to a real integer, $l$ [30].

## 2.2 Potential

The potential presented below and used here was originally developed to investigate Regge trajectories for both singular and non-singular potentials. Subsequently, it was applied successfully to the study of low-energy electron scattering from atoms leading to tenuously, weakly and strongly bound negative ion formation. Very recently, the robust potential employed here with Eq. (1) has been tested rigorously and applied successfully to electron scattering from the fullerene molecules $C_{20}$ to $C_{140}$. The extracted anionic BEs from the ground states TCSs for $C_{20}$ to $C_{92}$ matched excellently the measured EAs of these fullerenes, thereby giving great credence to the power of the Regge pole methodology to obtain credible results without the assistance of experiment and/or other theory. Also, as indicated in the introduction, the potential was used in the Mulholland formula to resolve the conundrum in the measured EAs for some of the lanthanide atoms. Here we carefully describe the essentials of the general potential in the context of the challenging fullerene molecules since the description of the complex heavy actinide atoms is similar to that of the fullerene molecules and the already investigated lanthanide atoms [17].

The representation of the complex heavy atomic, including the fullerene shell potential has continued to be a challenging and lingering problem that plagues conventional theoretical investigations of photon and electron interactions with fullerenes and complex heavy atoms. Recently, Baltenkov *et al* [31] investigated various popular $C_{60}$ shell model potentials and concluded that they generated nonphysical charge distributions. At the heart of these model potentials is the EA of the fullerene under investigation; it determines the parameters of the model potentials [32-46] and references therein. Furthermore, the rectangular potential was used as a fitting potential [47-50] with the parameters selected to best describe the experimental data [46, 50]. However, this procedure has been criticized [51]. Our potential, described below, because of its structure in the complex plane, has been used successfully for complex heavy systems such as atomic Au, Pt and the lanthanide atoms as well as the fullerene molecules.

For many years now theoretical calculations have been and are still struggling to go beyond the $C_{60}$ fullerene potential but without much success. Even the most recent investigation of electron scattering from A@$C_{60}$ (A= complex heavy atom, $C_{60}$ = fullerene molecule) modeled the $C_{60}$ cage by a standard attractive spherical potential [52]. Here the low-energy behavior of the elastic TCS is considered within the CAM representation of scattering using the simple approximation, *viz.* the Thomas–Fermi (T-F) theory, see for example [53, 54] and references therein. The exact form of the needed T-F potential q(r) for an atom/fullerene is determined from the universal T-F function χ(r) [55] (atomic units are used throughout)

$$q(r) = \frac{-Z \chi(r)}{r} \tag{2}$$

where Z is the atomic number of the target atom and the function χ(r) obeys the non-linear T-F differential equation

$$\frac{d^2 \chi(x)}{dx^2} = \frac{1}{\sqrt{x}} \chi^{3/2}(x) \tag{3}$$

Eq. (3) has the unusual boundary conditions:

$\chi(x) > 0, \quad x > 0$

$\chi(0) = 1$

$$\chi(x) \sim \frac{144}{x^3}, \quad x \to \infty \tag{4}$$

The variable x is given by x = r/μ, with μ = 0.8853$a_0$ $Z^{-1/3}$ and $a_0$ is the Bohr radius. A transcendental equation, Eq. (3) has only a numerical solution to date. A tabulation of the χ(x) function, also known as the universal T-F function, can be found in for example [56]. The Majorana solution of the T-F equation leads to a semi-analytical series solution [57]. In [58] the simple Padé approximant procedure was demonstrated to produce a remarkably good representation for the T-F exact solution.

An important accomplishment was achieved by Felfli et al [59] by replacing the exact function χ(x) by its rational function approximation in the Tietz context [60]. Within the T-F theory, Tietz used his potential [60] to investigate low-energy electron elastic scattering from many atoms, including complex heavy atoms and obtained analytical expressions for both the total cross sections and the scattering length. A similar potential to that of [60] was adopted in the investigation of resonances in low-energy electron scattering from neutral atoms [61]. In their investigation of Regge poles trajectories for non-singular potentials, Felfli et al [59] generated from the approximate function χ(x) a potential of the form, now known as the Avdonina-Belov-Felfli (ABF) potential,

$$U(r) = \frac{-Z}{r(1+\alpha Z^{1/3} r)(1+\beta Z^{2/3} r^2)} \tag{5}$$

where α and β are variational parameters. This potential has five turning points in the complex plane and the atomic/fullerene size, including deformation is also embedded therein through the presence of powers of Z as coefficients of r and $r^2$. Notably, our choice of the ABF potential, Eq. (5) is adequate as long as we limit our investigation to the near-threshold energy regime, where the elastic cross section is less sensitive to short-range interactions and is determined mostly by the polarization tail. Note also that the ABF potential has the appropriate asymptotic behavior, *viz.* $U(r) \sim -1/(\alpha\beta r^4)$ and accounts properly for the polarization interaction at low energies. The advantage of the replacement of the exact function χ(x) by its rational function approximation, namely the well-investigated Eq.(5), is that it is a good analytic function that can be continued to the complex plane.

In [62] Regge poles trajectories, viz. Im $\lambda_n(E)$ versus Re $\lambda_n(E)$ were investigated using the ABF potential $U(r)$, Eq. (5). The typical positions of all turning points and singularities of the potential, as well as all cuts in the complex plane are displayed in Fig. 1 of [62]. There are five turning points (TPs) and four poles of the ABF potential. Three of the TPs are located next to the poles (each pole is connected to the closely located TP by a cut) and the other two generate Regge poles (connected by a cut to each other). So there are a total of four cuts. The leading pole is located in the right half-plane only for sufficiently low energies; the energy region of interest in this paper. Furthermore, contrary to the case of singular potentials considered in [63], the Stokes' and anti-Stokes' lines topology for the ABF potential is characterized by a more complicated structure, thereby making the anti-Stokes' lines topology very difficult to implement [62].

The structure and topology in the complex plane of the ABF potential has been investigated extensively in the context of Regge poles trajectories [59, 62, 64, 65]. In particular, the physical Regge poles of the ABF potential, which are located in the right-hand CAM plane, were traced to the unphysical left-hand CAM plane as the impact energy increased indefinitely [64]. Resonant and non-resonant peaks, associated with the real part of a Regge pole trajectory, in electron-atom total scattering cross sections have been discussed recently [66]. One of the most important investigations of the importance of Regge poles trajectories in low-energy electron scattering using the ABF potential was carried out by Thylwe [27]. For atomic Xe the Dirac Relativistic and non-Relativistic Regge trajectories were contrasted near threshold and found to yield essentially the same Re $\lambda_n(E)$ when the Im $\lambda_n(E)$ was still very small, see Fig. 2 of [27]. This is a clear demonstration of the insignificant difference between the Relativistic and non-Relativistic calculations at low scattering energies, corresponding to possible electron attachment, leading to negative ion formation as resonances.

The effective potential
$$V(r) = U(r) + \lambda(\lambda+1)/2r^2 \tag{6}$$
is considered here as a continuous function of the variables *r* and complex *λ*. The potential, Eq. (5) has been used successfully with the appropriate values of *α* and *β*. When the TCS as a function of "*β*" has a resonance [26] corresponding to the formation of a stable long-lived negative ion, this resonance is longest lived for a given value of the energy which represents the ground state energy of the formed anion. This was found to be the case for all the systems we have investigated thus far and fixes the optimal value of "*β*" for Eq. (5); *α* has the optimal value of 0.2 throughout the paper.

For the numerical evaluation of the TCSs and the Mulholland partial cross sections, we solve the Schrödinger equation for complex values of L= *λ* and real, positive values of E

$$\psi'' + 2\left(E - \frac{L(L+1)}{2r^2} - U(r)\right)\psi = 0, \tag{7}$$

with the boundary conditions:
$$\psi(0) = 0,$$
$$\psi(r) \sim e^{+i\sqrt{2E}r}, \; r \to \infty. \tag{8}$$

We note that Eq. (8) defines a bound state when $k = \sqrt{2mE}$ is purely imaginary positive. We calculate the S-matrix, S(λ, k), poles positions and residues of Eq. (7) following a method similar to that of Burke and Tate [67]. In the method the two linearly independent solutions, $f_\lambda$ and $g_\lambda$, of the Schrödinger equation are evaluated as Bessel functions of complex order and the S-matrix, which is defined by the asymptotic boundary condition of the solution of the Schrödinger equation, is thus evaluated. Further details of the calculation may be found in [67].

Here we explain the effective use of the pole $\lambda$ (complex angular momentum) in the Mulholland formula, Eq. (1). From the pole $\lambda$, where Re$\lambda$(E) is an integer and Im$\lambda$(E) (= $\lambda_I$) → 0, we can determine from Eq. (1), according to the magnitude of Im$\lambda$(E), the shape resonances, long-lived metastable anions and ground state anionic resonances. The effective use of Im$\lambda$(E), $\lambda_I$ → 0 is demonstrated in the paper [26]. Although we have previously referred to Connor [68] for the physical interpretation of Im$\lambda$(E), the original interpretation was given by Regge himself [25]. The resonance width in energy is $\Gamma$ while $\lambda_I$ represents its width in angular momentum. The conjugate variable to energy is time, and the lifetime $\Delta t$ of the resonance satisfies the relation $\Delta t = 1/\Gamma$. Similarly, the conjugate variable in angular momentum is angle, and the angle $\Delta\theta$ through which the particle orbits during the course of the resonance satisfies the relation $\Delta\theta = 1/\lambda_I$. For a long-lived resonance, the lifetime $\lambda_I$ is small and $\Delta\theta$ is large, implying that the system orbits many times before decaying, while a large $\lambda_I$ value denotes a short lived state. For a true bound state, namely E < 0, Im$\lambda$(E) = 0 and therefore the angular life, 1/(Im$\lambda$(E)) → ∞, implying that the system can never decay. Obviously, in our calculations Im$\lambda$(E) is not identically zero, but small – this can be clearly seen in the figures: the dramatically sharp (long-lived) resonances hardly have a width as opposed to shape resonances for instance (see also [26] for comparison). We limit the calculations of the TCSs to the near-threshold energy region, namely below any excitation thresholds to avoid their effects.

### 3. Results.
### 3.1 Total cross sections

For a better understanding and appreciation of the physics involved in the calculations of the TCSs for the actinide atoms presented for the first time in Figs 2 through 6, we first present in Figs 1 left and right panels the known results for the structurally complicated Au atom and the $C_{60}$ fullerene molecule, respectively. For these systems the conventional theoretical methods experience considerable difficulties obtaining definitive EAs; in fact they need guidance. Generally, the low-energy electron scattering TCSs for complex heavy atoms such as Au and fullerene molecules such as $C_{60}$ are characterized by R-T minima, shape resonances (SRs) and dramatically sharp resonances, manifesting ground and metastable anionic formation, see also Figs 1.

Figures 1, left and right panels present respectively the TCSs for atomic Au, a typical complex heavy atom and for $C_{60}$, a standard fullerene molecule. Clearly seen is that the TCSs for these systems are characterized by ground and metastable negative ion formation. Their use here as examples is motivated by the fact that the energy positions, corresponding to the anionic BEs, of the ground state anionic resonance at 2.26 eV for Au, left panel and at 2.66 eV for $C_{60}$, right panel matched excellently the measured EAs of Au [13-15] and $C_{60}$ [69,70], see Table 1. This gives great credence to the ability of the Regge pole methodology to obtain reliable EAs for complex heavy atoms and fullerene molecules through the TCSs calculations. We first focus on the ground state TCS for Au (red curve). It is characterized by R-T minimum at about 0.7 eV, followed by a SR at about 1.0 eV. Subsequently, at the deep second R-T minimum at about 2.24 eV the Au negative ion is formed with the BE of 2.26 eV. The appearance of the two R-T minima in the TCS manifests that the polarization interaction has been accounted for adequately in the calculation [71]. The $C_{60}$ ground state TCS, Fig. 1(right panel), red curve behaves similarly to that of the Au atom, except that near threshold at about 0.065 eV the $C_{60}$ fullerene TCS exhibits a SR, manifesting the strong polarizability of the $C_{60}$ shell. As in the Au atom, the $C_{60}$ TCS exhibits two R-T minima as well. And the dramatically sharp resonance at 2.66 eV represents the $C_{60}$ anionic formation in the ground state. Of interest and importance in the ground state TCS of $C_{60}$ is the appearance near the ground state resonance of the metastable resonance at 1.86 eV. This could easily be mistaken for the BE of the $C_{60}$ ground state negative ion. Indeed, these results will certainly facilitate the understanding of the electron scattering TCSs for the actinide atoms being investigated here for the first

time. So, we identify and delineate the resonance structures in the TCSs of the being investigated actinide atoms and extract their ground state ionic BEs, yielding the sought after EAs.

The TCSs of Figs 2 through 6, rich in resonance structures, reflect essentially the sensitivity of these TCSs to the polarization interaction arising mainly from the complexes $5f^0 6d^2 7s^2$ for Th and $5f^n 6d^1 7s^2$ as n varies from n=2, 3, 4 and 5, corresponding respectively to the Pa, U, Np and Pu atoms. The sensitivity is manifested mainly through the BEs of the weakly bound excited anionic states. Notably, the ground state anionic BEs of these atoms, determining their EAs, are robust with respect to the transition from Th through Pu. A quick glance at the TCSs of Figs 2 through 6 shows the significant impact of the 6d orbital collapse to the 5f on the polarization interaction as we transition from Th $5f^0 6d^2 7s^2$ to Pa $5f^2 6d^1 7s^2$ as well as the size effect of Pu on its TCSs as we move from Np to Pu. The impact is clearly seen in the variation of the BEs of the first (highest) and the second anionic excited states, whose BEs are respectively 0.549 eV, 0.905 eV for Th; 0.395 eV, 0.874 eV for Pa and 0.220 eV, 0.507 eV for U. This behavior of the anionic BEs is similar to that of the lanthanide atoms already discussed in [72] and the fullerene molecules discussed in [3].

The electron scattering TCSs for ground, metastable and excited states of each actinide atom Th, Pa, U, Np and Pu are presented in the Figs 2, 3, 4, 5 and 6 respectively, while Table 1 summarizes the data for the investigated atoms, including the atomic Au and molecular $C_{60}$ for comparison. At a glance the TCSs for each atom in the figures appear to be complicated. However, these TCSs are readily understood and interpreted if we focus on a single color-coded curve at a time since each color-coded curve in each figure represents scattering from different states resulting in negative ion formation; a ground, metastable and higher excited states. Figure 2 contrasts the ground state TCS (red curve) with the metastable TCS (blue curve) and the higher excited states TCSs (green and pink curves) for the electron - Th scattering. Generally all the TCSs in the figure are characterized by R-T minima, SRs and dramatically sharp resonances corresponding to the Th⁻ anionic formation.

We briefly explain the physics behind the ground state TCS of Fig. 2; the explanation will be applicable to all the results presented in the figure. As the incident electron approaches the ground state Th atom closer, the atom becomes polarized, reaching maximum polarization manifested through the appearance of the first R-T minimum in the TCS at about 0.86 eV. With further increase in energy, the electron becomes trapped by the centrifugal potential whose effect is seen through the appearance of the SR at about 1.43 eV. As the electron leaks out of the barrier, the strong polarizability of the Th atom leads to the creation of the second deep R-T minimum at 3.10 eV. At the absolute minimum the long-lived ground state Th⁻ anion is formed; its BE is seen to be 3.09 eV. At this R-T minimum the Th atom is transparent to the incident electron and the electron becomes attached to it forming the stable ground state Th⁻ anion. The electron spends many angular rotations about the Th atom as it decays. The long angular life time of the ground state Th⁻ anion is determined by $1/(\text{Im } \lambda(E)) \to \infty$, since at the resonance Im $\lambda(E) \to 0$, see Eq. (1). Indeed, the appearance of the R-T minima in the electron scattering TCSs in Fig. 2, demonstrating the crucial importance of the polarization interaction, manifests that the polarization interaction has been accounted for adequately in the calculation, consistent with the conclusion in [71]. The metastable and excited states TCSs can be analyzed similarly; however, they are polarized differently as was discussed in [73] and have shorter life times as well, see [26] for further elaboration. The anionic BEs of the formed Th⁻ anions are summarized and compared with those of the other actinide atoms in Table 1. Also included in Table 1 are the calculated EAs from other theories.

Since the physics involved in the electron scattering TCSs calculations of the remaining actinide atoms Pa, U, Np and Pu is similar to that of the already discussed electron scattering calculation of the TCSs of the Th atom, here we will discuss some subtle differences in the calculated TCSs for these atoms when they are contrasted with one another. To this end, we first focus on the BEs of the highest excited anionic states (EXT-1 in Table 1). The 6d orbital collapse to the 5f in transitioning from Th ($5f^06d^27s^2$) to Pa ($5f^26d^17s^2$) impacts the polarization interaction reducing the anionic BE value of the first excited state (EXT-1) of Th from 0.549 to 0.395 eV for Pa. This value of 0.395 eV is further reduced to 0.220 eV in U ($5f^36d^17s^2$) since the effect of the 6d orbital collapse lingers on through U as was discussed for the lanthanide atoms in [72]. From U through Np to Pu the effect of the 6d orbital collapse has stabilized, demonstrated by the small variation in the first excited state anionic BEs of U, Np and Pu.

Next, we consider the effect of the 6d orbital collapse to the 5f on the second excited anionic states BEs (EXT-2, in Table 1). These are 0.905 eV and 0.874 eV for Th and Pa, respectively; as seen, the change is rather small. However, in U the value of 0.905 eV for Th has decreased significantly to 0.507 eV. Thereafter, the impact of the 6d orbital collapse on the BEs of Np and Pu remains small as it should. Another important impact of the 6d collapse is observed in the BE of the lowest metastable state, MS-1 in Table 1; it increases from 1.36 eV in Th to 1.49 eV in Pa. Thereafter, it stabilizes in U and Np. The size effect in Pu is significant; it creates an additional energy space thereby allowing for the creation of a new polarization-induced metastable TCS in Pu (brown curve) with a deep R-T minimum at about 0.13 eV and anionic BE value of 1.22 eV, see Fig. 6. Additionally, the widened energy space allows the lowest metastable BE value, MT-1 to increase to the larger value of 1.57 eV which is higher than those of U and Np. The increased energy space in Pu also impacts the ground state anionic BE of Pu, increasing it to 3.25 eV, which is larger than that of Np by 0.2 eV. The manifestation of the size dependence in the TCSs of Pu is similar to that found in the TCSs for the fullerene molecules [3]. Indeed, these results demonstrate the importance of identifying and delineating the resonance structures in the low-energy electron collision TCSs for complex heavy systems such as the actinides. We believe that these results will lead to a better understanding of low-energy electron scattering from complex heavy systems, leading to negative ion formation as well as to reliable determination of the EAs

Worth noting here is the behavior of the TCSs for the highest excited states, beginning from atomic U through Pu. Namely, the ratio of the second to the first R-T minima in the TCSs of U, Np and Pu is less than unity, contrary to that for the Th and Pa atoms. This behavior is typical of fullerene molecular behavior [3]. However, for fullerene molecular behavior, the additional requirement is that the near threshold ground state TCSs exhibit shape resonances. In none of these actinide atomic TCSs is this observed; all the ground state TCSs for these atoms U, Np and Pu exhibit atomic behavior [3]. The implication of this behavior requires careful investigation.

### 3.2 Comparison between our anionic BEs and existing theoretical EA values for the Th, Pa, U, Np and Pu atoms

Since Regge poles are generalized bound states, the extracted anionic BEs (ground, metastable and excited states) from the calculated TCSs should be reliable if the essential physics has been taken care of as in our case. As pointed out in the introduction, there are no measured EA values or TCSs for the actinide atoms available to compare with. However, for the actinide atoms considered here in this paper, theoretical EAs are available. Guo and Whitehead [23] used the generalized exchange local-spin-density functional theory combined with a relativistic correction (QR-LSD-GX-SIC theory with correlation-correction potential) to predict the EAs of the actinide elements. They concluded from their calculated

EAs that the Th⁻ anion is the most stable negative ion among the investigated actinides in agreement with the estimation [74]. Relativistic configuration-interaction (RCI) calculations were also used to obtain the EA values of Th, Pa, U, Np and Pu. Both the negative ions Ac⁻ [75] and Th⁻ [76] were found to form through the attachment of 7p and 6d electrons.

For a better understanding and appreciation of the comparison between our Regge pole calculated anionic BEs and the RCI calculated EAs for the actinide atoms displayed in Table 1, it is important to clarify the meaning of the EAs in the RCI calculations. In the works of Beck and collaborators [21,22,76,77] the EAs of the actinide atoms were generally *defined in terms of the attachment of a 7p electron to the neutral actinide atoms* forming stable negative ions. Consequently, the meaning of the EAs in the calculations of Beck and collaborators [21, 22, 76, 77] is strictly in this context. In our Regge pole calculated TCSs for the actinide and other complex heavy systems we obtained from the energy positions of the various very sharp peaks in the TCSs, representing negative ion formation, the anionic BEs. In our previous calculations of the electron scattering TCSs for the complex heavy Au and Pt atoms as well as the fullerene molecules, to name a few, we found that the BEs of the ground state negative ions, occurring at the second R-T minima of the ground state TCSs, matched excellently the measured EAs for the complex systems.

In Table 1 the comparison is made between our calculated anionic BEs for the ground states, GRS, the first and the second metastable states, MS-1 and MS-2, respectively, and the first and the second excited states, EXT-1 and EXT-2, respectively with the EAs calculated in [77] and those of Guo and Whitehead [23]. We can make general statements about the results. Firstly, the O'Malley and Beck [77] and the Guo and Whitehead [23] EAs do not agree. Secondly, our first (highest) anionic BEs are close to the EAs of O'Malley and Beck [77]; the best agreement being found for Pa (0.395 eV, our BE versus 0.384 eV, the EA of [77]). Thirdly, our anionic BEs for the second excited states agree reasonably well with the EAs of Guo and Whitehead for Th and Pa. However, for U, Np and Pu, the agreement in magnitude is excellent. The EA value of 1.17 eV for Th [23] and our second excited state BE value of 0.905 eV agree reasonably well and, supports the conclusion that the Th atom is the most stable actinide atom within the context of the calculations. For Pa the EA value of 0.55 eV [23] compares with our BE value of 0.875 eV. The difference in the results for Pa is attributed mainly to the effect of the 6d orbital collapse to the 5f impacting the polarization interaction significantly as we transition from Th $5f^06d^27s^2$ to Pa $5f^26d^17s^2$; the impact of this 6d orbital collapse is fully realized in U. We believe that the significant difference between our anionic BE value and the Guo and Whitehead EA value for Th is due mainly to the strong polarizability of the 6d orbital in Th $5f^06d^27s^2$ impacting strongly the polarization interaction, see the Th TCSs in Fig. 2. It is not clear to us whether the authors [23] accounted adequately for the important polarization interaction in their calculation.

In the comparison between our anionic BEs and the calculated EA values of the Beck *et al* group the extent of agreement between our anionic BEs and the EA values varies from atom to atom. For the Th atom the values are 0.549 eV (our BE) and 0.368 eV the EA of [77]; these values are reasonably close together. However, for Pu, they are 0.225 eV (our BE) versus 0.085 eV the EA of [77]; here a factor of about 2.6 difference is observed. The best agreement between our anionic BE value of 0.395 eV versus 0.384 eV [77] is obtained for the Pa atom. For the U and Np atoms the differences between our anionic BEs and the RCI EAs are factors of about 1.7 and 1.3, respectively. It is worth noting here that our BE value of 0.220 eV agrees much better with the previous EA value of 0.175 eV for U [22] than with the most recent EA value of 0.373 eV [77]. Over the years the EA values of the Beck group [21,22,76,77] have varied significantly; thus making it difficult to identify the possible source of the disagreement. This is indicative of the problems facing the standard theoretical methods in obtaining definitive EAs of complex heavy systems. Indeed, these sophisticated theoretical methods do not even seem to touch the problem of obtaining the reliable EAs. We believe that the source of these differences is the same as identified above in the comparison between the Regge pole calculated anionic BEs and the Guo and Whitehead [23] EA values. These comparisons, notwithstanding, the major objective of our paper is the investigation of ground, metastable and excited states negative ion formation in low-energy electron scattering from the considered complex actinide atoms and the attendant extraction of the anionic BEs.

Here a remark is appropriate regarding the importance of relativistic effects in the calculations using the Regge pole methodology of low-energy electron scattering TCSs for the complex heavy atoms, leading to stable negative ion formation. One of the most important investigations concerning the role of relativity in low-energy electron scattering from complex heavy atoms using the ABF potential was carried out by Thylwe [27]. For atomic Xe the Dirac Relativistic and non-Relativistic Regge trajectories were contrasted near threshold and found to yield essentially the same Re $\lambda(E)$ when the Im $\lambda(E)$ was still very small, see Fig. 2 of [27]. This is a clear demonstration of the insignificant difference between the Relativistic and non-Relativistic calculations at low-energy electron scattering, corresponding to possible electron attachment, leading to stable negative ion formation. This explains the great success of the Regge pole methodology in the present and previous investigations of low-energy electron scattering from complex heavy atoms including fullerenes molecules, leading to stable negative ion formation. The extraction from the calculated TCSs of the anionic ground state BEs that match excellently the measured EAs is the Regge pole methodology's great accomplishment. Indeed, the methodology requires no assistance from either experiment or other theory to achieve the remarkable results for complex heavy systems.

## 4. Discussion and Conclusion

We have used our robust Regge pole methodology to investigate low-energy electron scattering from the actinide atoms Th, Pa, U, Np and Pu through the elastic TCSs calculations, leading to negative ion formation. We found that the TCSs for these actinide atoms are characterized by ground, metastable and excited negative ion formation. The extracted BEs of the resultant negative ions have been employed to understand the calculated EAs of the actinide atoms obtained using the RCI methodology [21,22,76,77] and those calculated by Guo and Whitehead [23], with the conclusion that the EAs of [21,22,76,77] correspond essentially to the BEs of the highest excited anionic states, EXT-1 in Table 1. The calculated EAs of Guo and Whitehead on the other hand are identified with the BEs of the second excited anionic states, EXT-2 in Table 1 of the formed negative ions during the collisions. Our extracted ground state anionic BEs, located at the second R-T minima of the TCSs, indicated as very sharp resonances in the Figs. 2 through 6 should be taken as the EAs of the investigated actinide atoms consistent with that done for the TCSs of the Eu, Au and Pt atoms and the fullerene molecules to name a few systems.

Importantly, the ground state negative ions are formed at the second R-T minima of the electron scattering TCSs during the electron collisions with the ground state actinide atoms. This is where the maximum polarization of the systems occurs. And this could be useful in guiding future theoretical and experimental investigations of the EAs of complex heavy systems, such as the actinide atoms. Once the reliable anionic BEs of the complex systems have been determined as in our case here, then the sophisticated theoretical calculations can be employed for the fine-structure splitting determination as well as the spin-orbit coupling effects. Thylwe [27] demonstrated that when the Im λ(E) was still very small, near threshold the Relativistic and non-Relativistic Regge trajectories yielded essentially the same Re λ(E), reflecting the insignificant contribution of relativity to the TCSs calculations provided that the essential physics has been accounted for. Also the closeness of some of our BEs of the highest excited anionic states to the RCI EAs and our BEs of the second excited anionic states to the EA values of Guo and Whitehead [23] also support the above conclusion, even though the theoretical EAs above correspond to the anionic BEs for excited states. We emphasize here that our anionic BEs are robust and firm.

In conclusion, our TCSs for the actinide atoms Th, Pa, U, Np and Pu, rich in ground, metastable and excited states negative ion formation, yield the anionic ground states BEs whose values are 3.09 eV, 2.98 eV, 3.03 eV, 3.06 eV and 3.25 eV for Th, Pa, U, Np and Pu, respectively. These should be considered as the EAs of the investigated actinide atoms and suggest the use of low-energy electron scattering for determining unambiguously the EAs of complex heavy systems such as the actinide atoms. Indeed, the increasing importance of the polarization interaction with the size of the actinide atom is manifested in the Pu TCSs through the appearance of the additional polarization-induced metastable TCS, brown curve in Fig. 6, with the deep R-T minimum and the anionic BE value of 1.22 eV. Furthermore, our calculated TCSs will also be important and useful in the understanding of the roles of the outer 6d and 5f subshell electrons in chemical reactions involving actinide atoms through the negative ion formation, as well as in nanocatalysis. Additionally, the exploration of negative ion formation in low-energy electron scattering from complex heavy atoms and empty fullerene molecules is expected to lead to a better understanding and interpretation of the resonance structures observed in electron scattering and photoionization cross sections of the endohedral A@$C_n$ ($C_n$ = fullerene molecule) systems.

**Acknowledgements**
This research was supported by the US DOE, Division of Chemical Sciences, Geosciences and Biosciences, Office of Basic Energy Sciences, Office of Energy Research. The computing facilities of the National Energy Research Scientific Computing Center are greatly appreciated.

**Table 1:** Negative ions binding energies (BEs), in eV and energy positions of Ramsauer-Townsend (R-T) minima, in eV obtained from the TCSs for the atoms Au, Th, Pa, U, Np and Pu. GRS, MS-$n$ and EXT-$n$ represent respectively ground, metastable and excited states. The experimental EAs, EXPT and the theoretical EAs, RCI [77] and GW [23] are also presented.

| System | BEs GRS | BEs MS-1 | BEs MS-2 | EAs EXPT | BEs EXT-1 | BEs EXT-2 | R-T GRS | R-T MS-1 | BEs/EAs Theory | EAs RCI[77] | EAs GW[23] |
|---|---|---|---|---|---|---|---|---|---|---|---|
| Au | 2.26 | 0.832 | - | 2.301[14] 2.306[15] | 0.326 | - | 2.24 | - | 2.26[78] | - | - |
| $C_{60}$ | 2.66 | 1.86 | 1.23 | 2.684[69] 2.666[70] | 0.203 | 0.378 | 2.67 | 1.86 | 2.66[4] | - | - |
| Th | 3.09 | 1.36 | - | - | 0.549 | 0.905 | 3.10 | 1.38 | 0.368 [76] | 0.368 | 1.17 |
| Pa | 2.98 | 1.49 | - | - | 0.395 | 0.874 | 3.00 | 1.50 | 0.222[21] | 0.384 | 0.55 |
| U | 3.03 | 1.44 | - | - | 0.220 | 0.507 | 3.01 | 1.43 | 0.175[22] | 0.373 | 0.53 |
| Np | 3.06 | 1.47 | - | - | 0.248 | 0.521 | 3.05 | 1.49 | - | 0.313 | 0.48 |
| Pu | 3.25 | 1.57 | 1.22 | - | 0.225 | 0.527 | 3.23 | 1.58 | - | 0.085 | -0.50 |

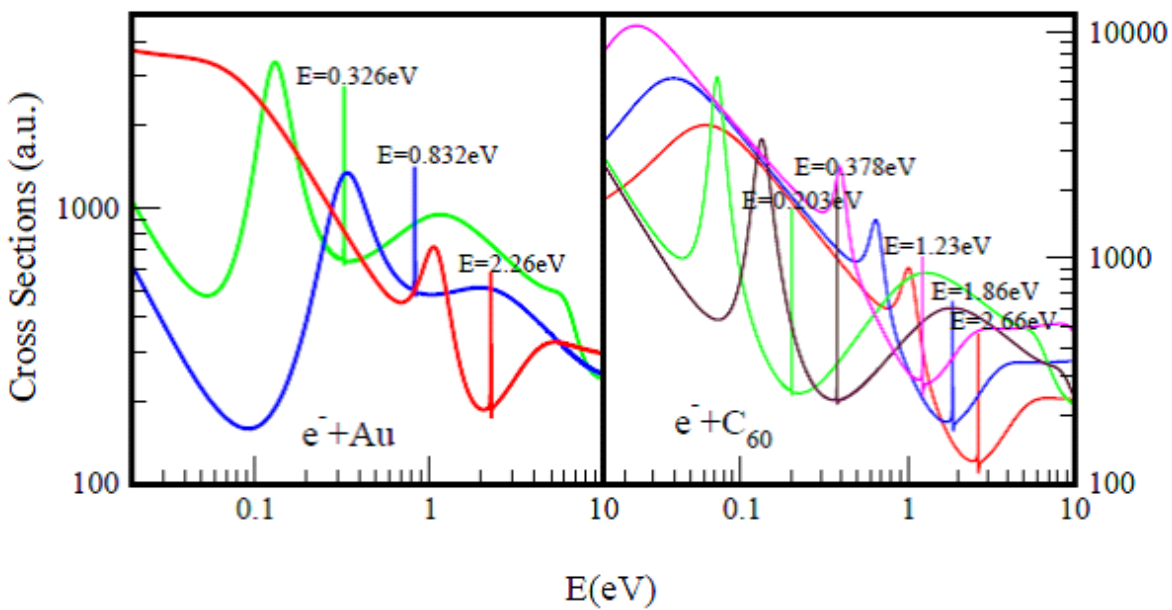

**Figure 1:** Total cross sections (a.u.) for electron elastic scattering from atomic Au (left panel) and the fullerene molecule $C_{60}$ (right panel) are contrasted. For atomic Au the red, blue and green curves represent TCSs for the ground, metastable and excited states, respectively. For the $C_{60}$ fullerene the red, blue and pink curves represent TCSs for the ground and the metastable states, respectively while the brown and green curves denote TCSs for the excited states. The dramatically sharp resonances in both figures correspond to the $Au^-$ and $C_{60}^-$ negative ions formation during the collisions.

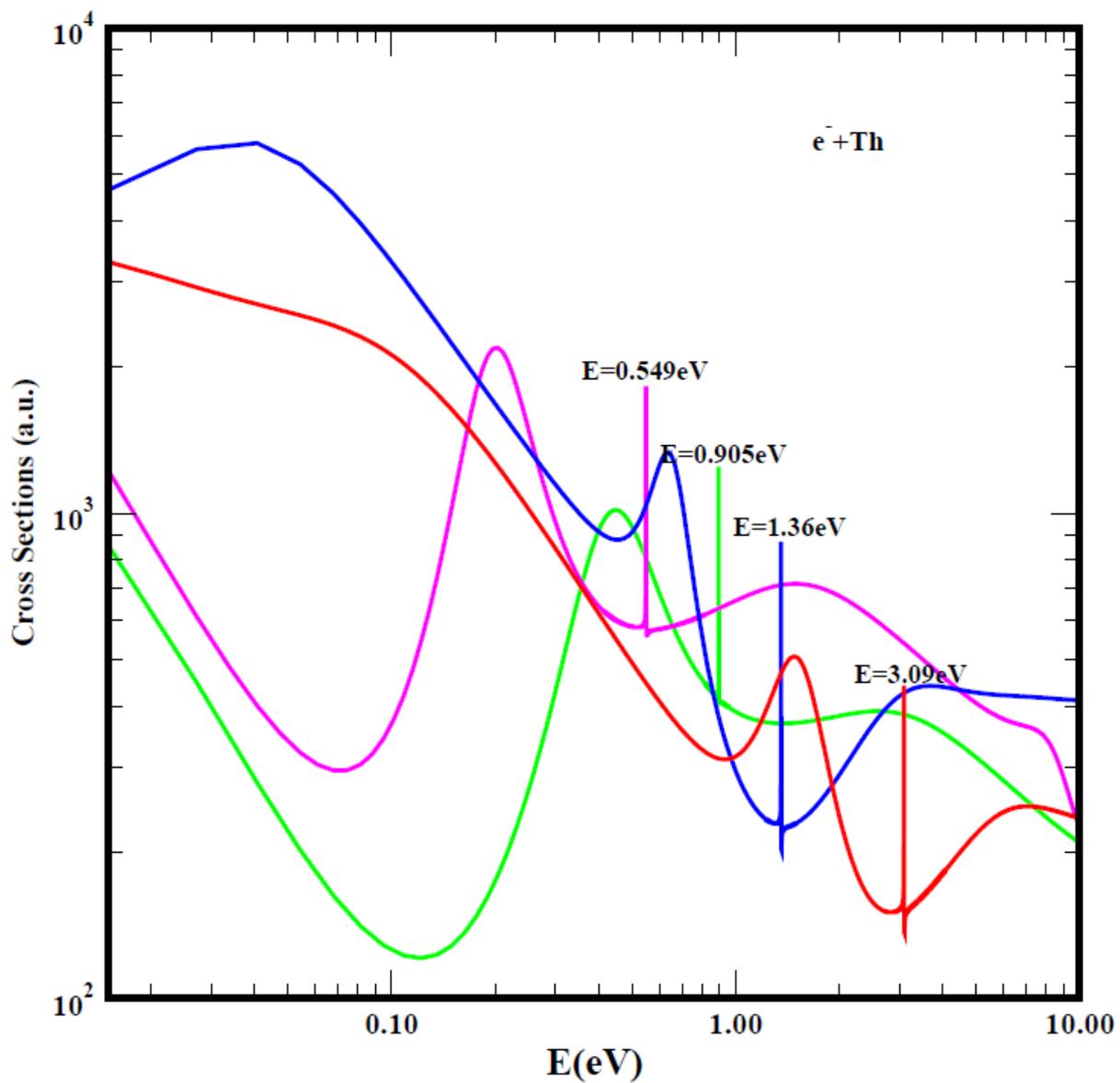

**Figure 2:** Total cross sections (a.u.) for electron elastic scattering from atomic Th. The red and blue curves represent TCSs for the ground and the metastable states, respectively. The pink and green curves stand for TCSs for the excited states. The dramatically sharp resonances correspond to the Th⁻ anions formation during the collisions

08/24/201808/24/2018

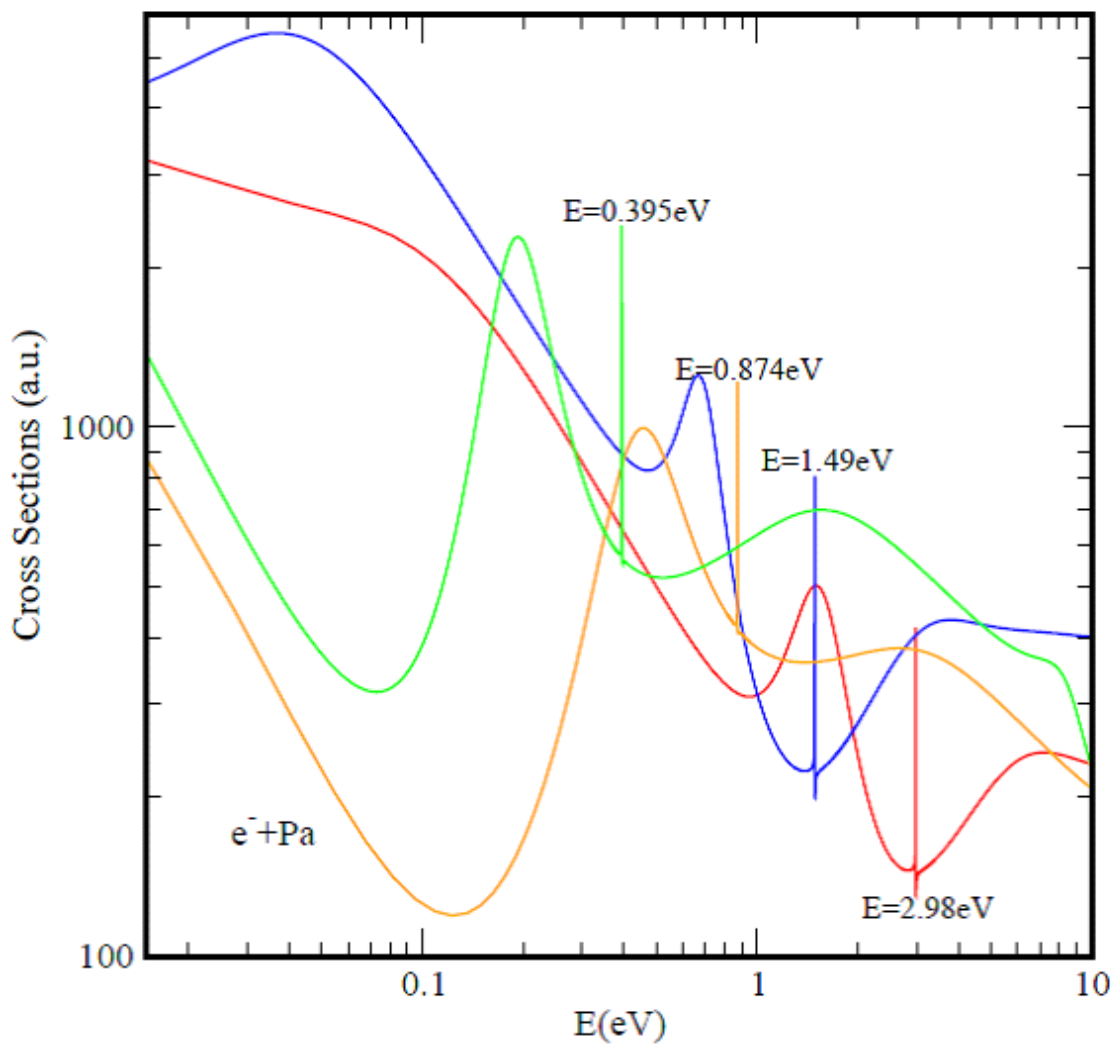

**Figure 3:** Total cross sections (a.u.) for electron elastic scattering from atomic Pa. The red and blue curves represent TCSs for the ground and the metastable states, respectively. The orange and green curves represent TCSs for the excited states. The dramatically sharp resonances correspond to the Pa⁻ anions formation during the collisions.

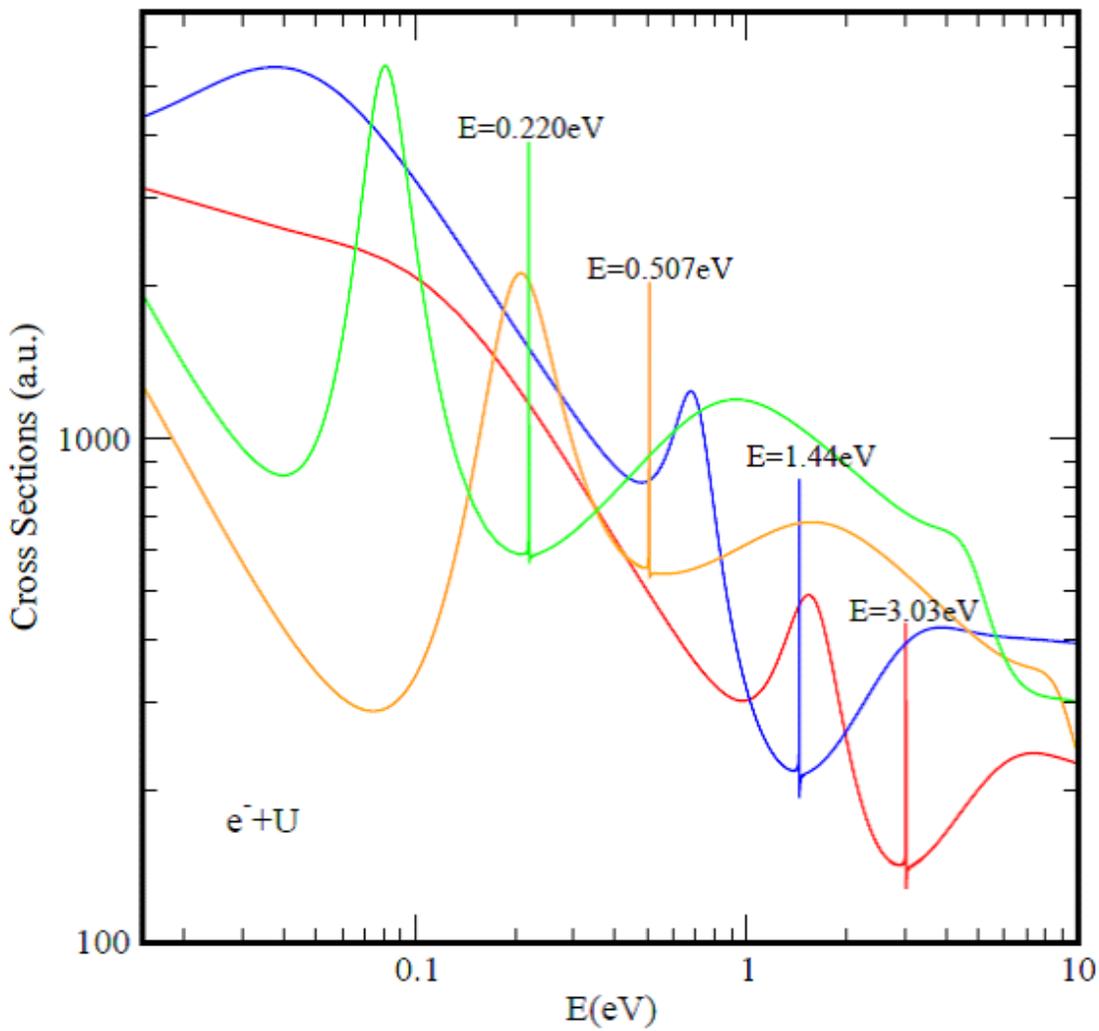

**Figure 4:** Total cross sections (a.u.) for electron elastic scattering from atomic U. The red and blue curves represent TCSs for the ground and the metastable states, respectively. The orange and green curves represent TCSs for the excited states. The dramatically sharp resonances correspond to the U⁻ anions formation during the collisions

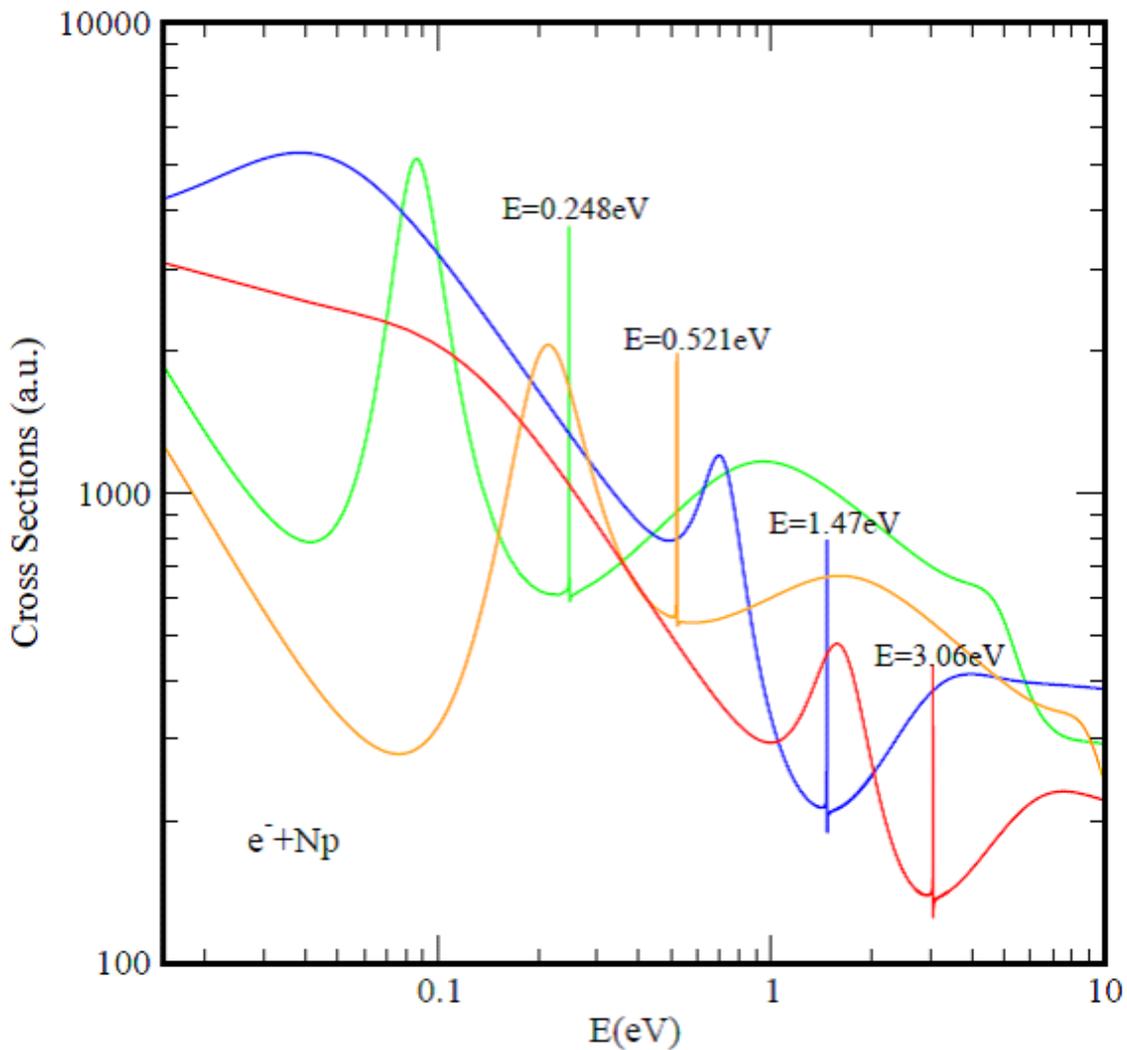

Figure 5: Total cross sections (a.u.) for electron elastic scattering from atomic Np. The red and blue curves represent TCSs for the ground and the metastable states, respectively. The orange and green curves represent TCSs for the excited states. The dramatically sharp resonances correspond to the Np⁻ anions formation during the collisions.

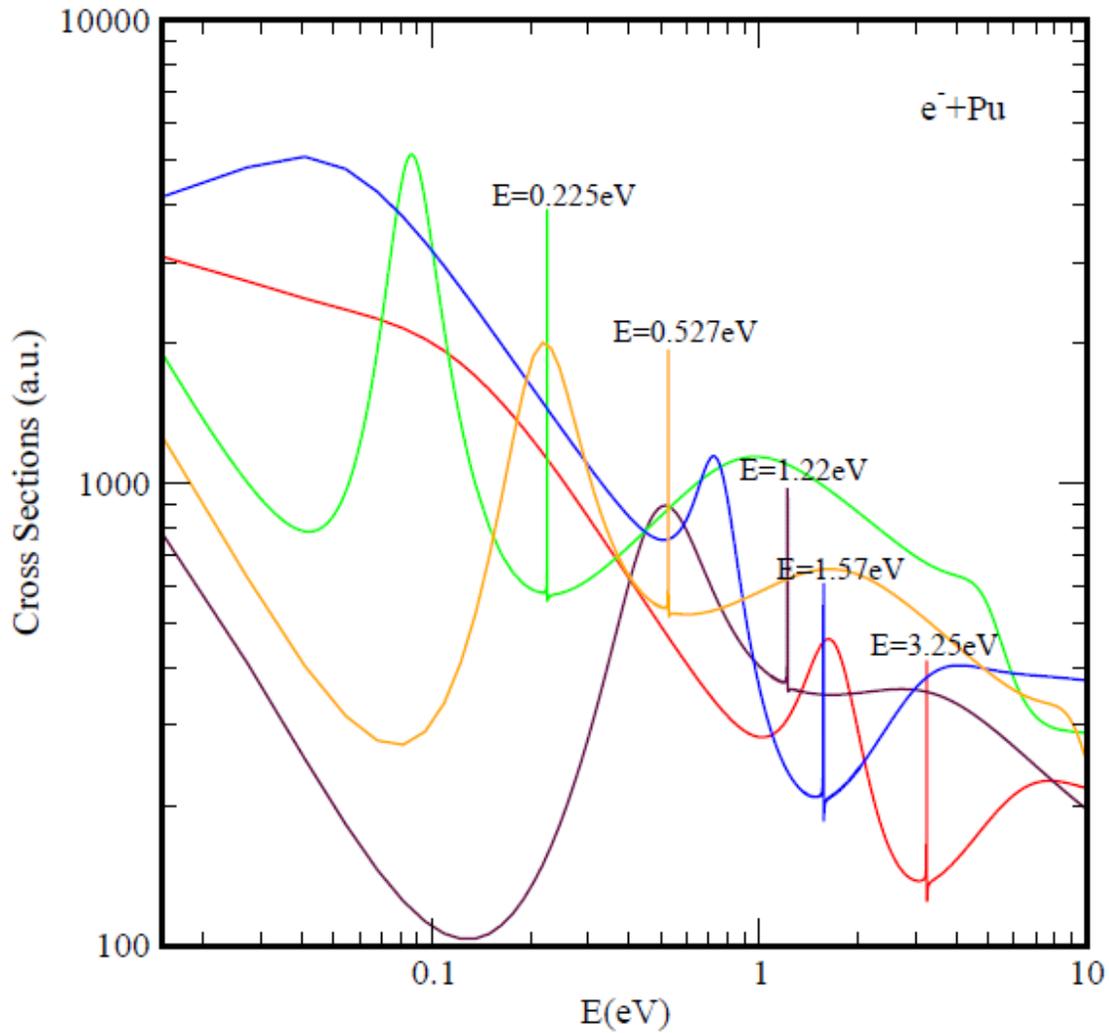

**Figure 6:** Total cross sections (a.u.) for electron elastic scattering from atomic Pu. The red and the blue and brown curves represent TCSs for the ground state and the metastable states, respectively. The orange and the green curves denote TCSs for the excited states. The dramatically sharp resonances correspond to the Pu⁻ anions formation during the collisions